\documentclass[twocolumn,showpacs,preprintnumbers,amsmath,amssymb]{revtex4}

\usepackage{graphicx}
\usepackage{dcolumn}
\usepackage{bm}
\usepackage{natbib}

\begin{document}

\title{Quantifying fusion born ion populations in magnetically confined plasmas using ion cyclotron emission}

\author{L Carbajal$^{1,2}$ \let\thefootnote\relax\footnote{This submission was written by the author(s) acting in (his/her/their) own independent capacity and not on behalf of UT-Battelle, LLC, or its affiliates or successors.}}%
 \email{l.carbajal-gomez@warwick.ac.uk.}
 \author{R O Dendy$^{2,3}$}
 \author{S C Chapman$^{2}$}
 \author{J W S Cook$^{4,2}$}
 \affiliation{$^{1}$ Oak Ridge National Laboratory, PO Box 2008, Oak Ridge, TN 37831-6169 USA}

 \affiliation{$^{2}$Centre for Fusion, Space and Astrophysics, Department of Physics, Warwick University, Coventry CV4 7AL, UK}

\affiliation{$^{3}$CCFE, Culham Science Centre\\ Abingdon, Oxfordshire OX14 3DB, UK}

\affiliation{$^{4}$ First Light Fusion Ltd, Oxford Industrial Park, Yarnton, Oxfordshire OX5 1QU}

\begin{abstract}
Ion cyclotron emission (ICE) offers unique promise as a diagnostic of the fusion born alpha-particle population in magnetically confined plasmas. Pioneering observations from JET and TFTR found that ICE intensity $P_{ICE}$ scales approximately linearly with the measured neutron flux from fusion reactions, and with the inferred concentration, $n_\alpha/n_i$, of fusion-born alpha-particles confined within the plasma. We present fully nonlinear self-consistent kinetic simulations that reproduce this scaling for the first time. This resolves a longstanding question in the physics of fusion alpha particle confinement and stability in MCF plasmas. It confirms the magnetoacoustic cyclotron instability (MCI) as the likely emission mechanism and greatly strengthens the basis for diagnostic exploitation of ICE in future burning plasmas.

\end{abstract}

\maketitle
The containment of populations of alpha-particles, born in nuclear fusion reactions between thermal ions, is central to the exploitation of magnetically confined plasmas for energy generation. In contrast to the thermal (10 keV to 20 keV) deuterium and tritium ions which form the bulk of a burning plasma, alpha-particles are born at $3.5MeV$ and in consequence are only marginally confined even in the largest contemporary tokamak experiment, JET. These alpha-particles must remain confined while they transfer energy, via collisions with electrons, to the bulk plasma on a slowing-down timescale $\sim 0.5s$ which is of order $10^7 $ gyro-periods. The alpha-particles also couple non-collisionally to the thermal plasma and magnetic field, for example by fast timescale collective resonant interaction with MHD modes. A key objective of  ITER is to explore the confinement physics of fusion-born alpha-particles under conditions where, unlike at present, they are the primary heating source for the plasma. Direct observation of the spatial distribution and collective dynamics of this alpha-particle population is challenging: they are only $\sim 0.1$ \% of the plasma particles, their single-particle radiation is in the background, and in-situ probe measurements are impossible.

Ion cyclotron emission (ICE) offers unique promise as a diagnostic of the alpha-particles, and has been proposed for this role in future deuterium-tritium plasmas in JET and ITER \cite{McClements2015}. Pioneering observations from JET and TFTR \cite{Cottrell1993,X3,Dendy1995} found that ICE intensity $P_{ICE}$ scales approximately linearly with the measured neutron flux from fusion reactions, and with the inferred concentration, $n_\alpha/n_i$, of fusion-born alpha-particles confined within the plasma. Understanding the physical basis for this scaling is important because, in future experiments with burning plasma, measurements of the observed ICE intensity could then directly diagnose the spatial distribution and temporal evolution of the confined alpha-particle population. From its suprathermal nature, it is evident that ICE arises from a collective radiative instability, and the leading candidate is the magnetoacoustic cyclotron instability (MCI) \cite{X4b,X4a,McClements1996,X4c,X5a,X5b}. The MCI is driven by the free energy of the alpha-particles, whose distribution in velocity-space can exhibit a natural population inversion, localised in space at the outer mid-plane edge of the plasma.  The way in which MCI saturation physics leads to the observed scaling of $P_{ICE}$ with $n_\alpha/n_i$ has not hitherto been determined. In particular, the lower cyclotron harmonics ($1 \leq l \leq 6$) are typically linearly stable against the MCI, hence their existence and magnitude is solely determined by nonlinear physics \cite{Carbajal2014,Cook}.
We have therefore performed kinetic numerical simulations of the MCI which are fully nonlinear and incorporate self-consistent particle dynamics and field evolution. Identifying the saturated intensity of the simulated MCI-excited fields with $P_{ICE}$,  we are able to reproduce the observed approximately linear scaling of $P_{ICE}$ with $n_\alpha/n_i$.
These first-principles simulations also enable us to directly identify the underlying physics of the observed scaling.

In the early JET D-T experiments \cite{X2a,Cottrell1993,Dendy1995,X2b}, high power NBI was used to heat the plasma. By comparing plasmas where only deuterium was used as the NBI source to those where a mixture of deuterium and tritium was used, a linear correlation was identified between the measured ICE power at the second cyclotron harmonic $\omega = 2\Omega_\alpha$ and the measured fusion reactivity, for experiments spanning a factor of a million in the neutron source rate (cf. Fig.~3 of Ref.~[\onlinecite{Dendy1995}]). In addition, time resolved measurements during specific JET D-T plasmas show linear scaling of ICE intensity at $\omega = 2\Omega_\alpha$ with local edge alpha-particle concentration as time evolved (cf. Fig.~4 of Ref.~[\onlinecite{Dendy1995}]). These results strongly suggest that the source of ICE in those plasmas was the fusion-born alpha-particles resulting from D-T fusion reactions. Furthermore, in all these large tokamak plasmas, the ICE power spectrum displayed peaks at sequential cyclotron harmonics of alpha-particles at the outer mid-plane edge. These characteristics are largely replicated by recent second generation ICE measurements, to which the work reported here is also relevant.
ICE is used as a diagnostic for energetic ions lost from DIII-D tokamak plasmas due to MHD activity \cite{15}. ICE is observed from ASDEX-Upgrade \cite{16,17} and JT-60U \cite{18,19} tokamak plasmas, with spectral peaks at edge cyclotron harmonics of ion species that include the energetic products of fusion reactions in pure deuterium, namely protons, tritons, and helium-3 ions. ICE from sub-Alfv\'enic beam ions, injected at tens of keV to heat the plasma, was observed from TFTR tokamak plasmas, and is also interpreted in terms of the MCI \cite{20}. ICE which is probably of this kind is also used as a diagnostic of lost fast ions in large stellarator-heliotron plasmas in LHD \cite{21}. Very recently, ICE with spectral peak separation equal to the edge proton cyclotron frequency has been reported from deuterium plasmas in the KSTAR tokamak \cite{21a}.

Radiation similar to ICE is also observed in space plasmas, driven by energetic ions with drifting ring-like velocity distributions, and has also been interpreted in terms of the MCI \cite{McClements1993,McClementsJGR}. Its nonlinear development can be resolved in  fully kinetic simulations, for example in reforming perpendicular shocks in space plasmas which show clear spectral peaks at consecutive ion cyclotron harmonics, suggesting the occurrence of ICE in the region behind the shock front \cite{Rekaa}. Fully self consistent kinetic simulations of the MCI may thus be relevant to ICE occurring in space as well as fusion plasmas \cite{Cook, Carbajal2014}. ICE phenomenology could also be exploited for ``alpha channeling'' in MCF plasmas \cite{X2c}.

 In this Letter we simulate the MCI in the hybrid approximation  in which the ion dynamics (both energetic alpha-particles and majority thermal ions) follows the full Lorentz force, and the electrons are represented as a fluid. Our simulations self-consistently evolve the low frequency Maxwell-Lorentz system for all three vector components of the particle velocities and electric and magnetic fields, in one spatial dimension. We study the MCI for concentrations of alpha-particles relative to thermal ions that range between $n_\alpha/n_i=10^{-4}$ and $10^{-3}$. The lower value is the smallest that is realizable, with reasonable computational resources, while carrying the MCI into its nonlinear regime.

We demonstrate that the physical basis for the observed approximately linear scaling of $P_{ICE}$ with $n_\alpha/ n_i$ is as follows. The most strongly linearly unstable cyclotron harmonic modes $\omega = l \Omega_\alpha$ ($8 \leq l \leq 10$) grow rapidly, initially at their linear growth rates which scale as  $\sqrt n_\alpha$. This growth is so rapid that its linear phase terminates after only a few gyro-periods. Beating between these linearly excited modes drives the growth of the lower harmonic ($1 \leq l \leq 6$) linearly stable modes. Our fully nonlinear simulations show that these energy flows are such that, at saturation, all mode intensities scale approximately with the concentration of alpha-particles.

\begin{figure}
\begin{center}
\includegraphics[scale=0.6]{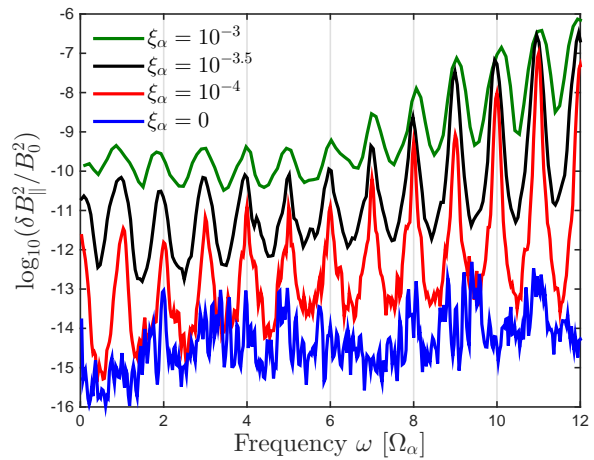}
\end{center}
\caption{Log-linear plot of the  power spectral density of the normalised perturbed magnetic field $\delta B_\parallel/B_0$ in the fully nonlinear regime of the simulations at increasing concentrations $\xi_\alpha = n_\alpha/n_i$ of energetic particles (red, black, green solid lines). The noise level in the simulations is indicated by the bottom trace(blue solid line), which is a simulation for background plasma only, $\xi_\alpha =0$. The values $\xi_\alpha$ for the three
simulations that include energetic particles
are a half decade apart, and the corresponding heights of the peaks in $\log_{10}\left( \delta B^2_\parallel/B^2_0 \right)$ can be seen to increase approximately linearly between these traces for harmonics $\leq 6$.}
\label{Fig1}
\end{figure}

\begin{figure}
\begin{center}
\includegraphics[scale=0.51]{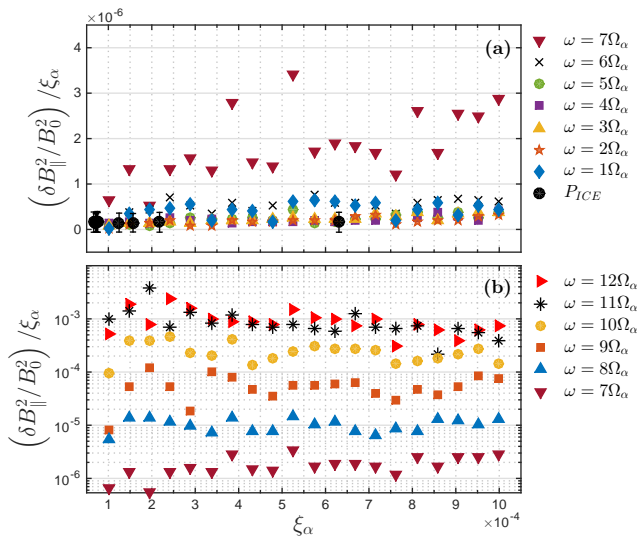}
\end{center}
\caption{The perturbed normalized magnetic field energy in the fully nonlinear regime of the simulations $\delta B_\parallel^2/B_0^2$ at the peak of each cyclotron harmonic, divided by concentration of energetic ions $\xi_\alpha$, is plotted versus $\xi_\alpha$. The linearly stable modes $\omega = \Omega_\alpha$ to $6\Omega_\alpha$ and the marginally  unstable mode $\omega = 7 \Omega_\alpha$ are plotted in upper panel (a).
 The black filled circles in this panel show the measured ICE power $P_{\mbox{ICE}}$ at $\omega = 2 \Omega_\alpha$ inferred from Fig.~4 of Ref.~[\onlinecite{Dendy1995}]; the error bars of $P_{\mbox{ICE}}$ correspond to the uncertainty of $\pm 6$ dB in the detected ICE power reported in Ref.~[\onlinecite{Cottrell1993}]. Lower panel (b) plots, in the same format, the linearly unstable modes $\omega = 7-12 \Omega_\alpha$. Note the different scale of the ordinate in the two panels. On these compensated plots, a straight horizontal trend implies a relationship $\delta B_\parallel^2/B_0^2 \sim \xi_\alpha$.}
\label{Fig2}
\end{figure}

We use the PROMETHEUS++ code \cite{PROMETHEUS} which implements the hybrid approximation \cite{Gingell2012,Gingell2014} as applied to the MCI in Ref.~[\onlinecite{Carbajal2014}]. In this approximation, the plasma is quasineutral in this low-frequency (Darwin) limit, so that the total number density is $n = n_e = \sum Z_j n_j$, where $n_e$ is the electron number density, and $Z_j$ and $n_j$ are the atomic number and the number density of the species $j$.  In all our simulations, the direction of variation is along the $x$-axis, perpendicular to the background magnetic field, $\bm{B}_0 = B_0 \bm{z}$.
\begin{figure}
\begin{center}

\includegraphics[scale=0.51]{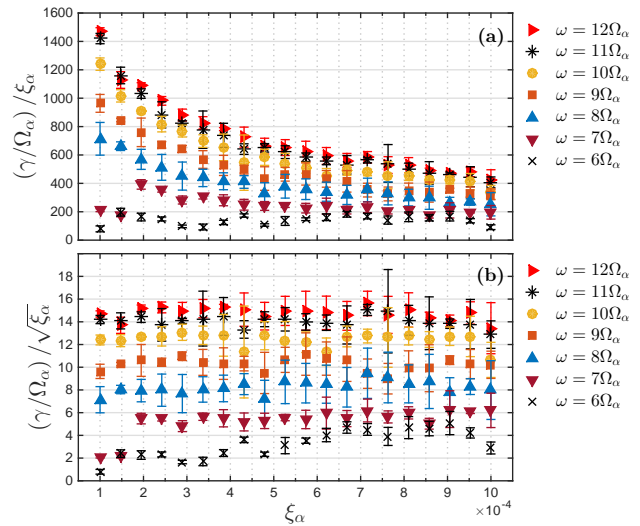}
\end{center}
\caption{The growth rate for the MCI obtained at early times in the simulation, $\gamma$, normalized to $\Omega_\alpha$ and divided by concentration of energetic ions $\xi_\alpha$ (upper panel (a)) and divided by $\sqrt \xi_\alpha$ (lower panel (b)), are plotted versus $\xi_\alpha$ for linearly stable mode $\omega = 6\Omega_\alpha$ and
linearly unstable modes $\omega = 7\Omega_\alpha$ to $12\Omega_\alpha$.
On these compensated plots, a straight horizontal trend implies a relationship $\gamma/\Omega_\alpha \sim \xi_\alpha$ (upper panel (a)) and $\gamma/\Omega_\alpha \sim \sqrt \xi_\alpha$ (lower panel (b)).}
\label{Fig3}
\end{figure}

The plasma parameters are similar to those of the D-T plasmas of JET at the outer mid-plane edge, \cite{Cottrell1993} with background magnetic field $B_0=2.1$T, total number density $n=1\times10^{19}$ m$^{-3}$ and electron temperature $T_e=1$keV. The background population of deuterons is in thermal equilibrium with the fluid electrons, and the local minority alpha-particle population $n_\alpha$ is modelled as a ring-like distribution in velocity space $f_\alpha(v_\parallel, v_\perp) = \delta (v_\parallel) \delta (v_\perp - u_\perp)$. Here $u_\perp$ is the magnitude of the initial perpendicular velocity of the alpha-particles, corresponding to the energy of 3.5 MeV at which alpha-particles are born in D-T fusion reactions. 
This experimentally motivated distribution function corresponds to a simplified version of the one used in the analytical theory of the MCI in Ref.~[\onlinecite{McClements1996}].

The number density is initially constant across the simulation domain, subject to noise fluctuations. As a check against sensitivity to simulation noise, we repeated all simulations  using both a uniform random number generator and a ``quiet start'' \cite{Sydora1999}. The latter method reduces the small fluctuations in the initial number density $n$, which can provide an extra source of free energy.
The periodic simulation domain consists of 8192 grid cells, each of length equal to $r_D/\sqrt{2}$, where $r_D = v_{T_D}/\Omega_D$ is the deuteron Larmor radius, $v_{T_D}^2 = k_B T_D/m_D$ is the deuteron thermal velocity, and $\Omega_D$ is the deuteron cyclotron frequency. In this way, we resolve in detail the gyro-motion of both ion species; alpha-particle dynamics are particularly highly resolved, given that $r_\alpha/r_D \approx 41$. The simulations use 2000 macro-particles per grid cell to represent each ion population.

To compare our hybrid simulations with measured ICE amplitude it is essential to access the fully nonlinear regime.
These hybrid simulations of the MCI then recover the observed scaling properties of ICE intensity $P_{ICE}$ with respect to the concentration of energetic ions $\xi_\alpha=n_\alpha / n_i$, observed in JET, giving further support to the MCI as the mechanism underlying ICE. We perform hybrid simulations of the MCI with twenty different values of $\xi_\alpha$, ranging from $\xi_\alpha = 10^{-4}$ to $\xi_\alpha = 10^{-3}$.
The first result drawn from these simulations is the excitation of perpendicular propagating fast Alfv\'en waves at consecutive ion cyclotron harmonics of alpha-particles, for all the values of $\xi_\alpha$.
 Power spectra of the normalised parallel perturbed magnetic field energy $\delta B_\parallel^2/B_0^2$ for $\xi_\alpha = 10^{-4}$,  $10^{-3.5}$, and $10^{-3}$, respectively are shown in Fig.~\ref{Fig1}.
 Noise is reduced in these simulations using a quiet start \cite{Sydora1999}, and  the spectral peaks at the cyclotron harmonics  emerge well above the noise level. The values chosen for $\xi_\alpha$ in Fig. 1 are logarithmically equally spaced. The corresponding spectral peak power for the nonlinearly unstable modes $1 \leq l \leq 6$  can also be seen to increase logarithmically - the relative heights of these peaks are linearly spaced on this log-linear plot. These MCI simulations thus suggest linear scaling of $P_{ICE}$ with $\xi_\alpha$, consistent with experiment.

  We now examine this in more detail. In Fig. 2 we plot versus $\xi_\alpha$
  the perturbed normalized magnetic field energy in the fully nonlinear regime of the simulations $\delta B_\parallel^2/B_0^2$ at the peak of each cyclotron harmonic, divided by concentration of energetic ions $\xi_\alpha$. On these plots, a straight horizontal trend implies a linear relationship $\delta B_\parallel^2/B_0^2 \sim \xi_\alpha$.
  The linearly stable modes $\omega = \Omega_\alpha$ to $6\Omega_\alpha$ and the marginally  unstable mode $\omega = 7 \Omega_\alpha$ are plotted in upper panel (a).
 The black filled circles in this panel show the measured ICE power $P_{\mbox{ICE}}$ at $\omega = 2 \Omega_\alpha$ inferred from Fig.~4 of Ref.~[\onlinecite{Dendy1995}]; the error bars of $P_{\mbox{ICE}}$ correspond to the uncertainty of $\pm 6$ dB in the detected ICE power reported in Ref.~[\onlinecite{Cottrell1993}].  We can see that there is good correspondence between the observations and the nonlinear peak power at $\omega = \Omega_\alpha$ to $6\Omega_\alpha$. The marginally linearly unstable mode at  $l=7$ is plotted in both panels (a) and (b) where the ordinate are on different scales. It marks the transition between the linearly stable modes $\omega = \Omega_\alpha$ to $6\Omega_\alpha$ and the linearly unstable modes $\omega = 7 \Omega_\alpha$ to $12\Omega_\alpha$. These linearly unstable modes exhibit approximately linear scaling of intensity with $\xi_\alpha$ but with significant scatter.
In Fig.3 we examine the behaviour earlier in the simulations, during the linear growth phase of the MCI of modes $\omega = 7 \Omega_\alpha$ to $12\Omega_\alpha$.
In this figure, the growth rate for the MCI obtained at early times in the simulation, $\gamma$, (normalized to $\Omega_\alpha$) is plotted in the upper panel (a) divided by concentration of energetic ions $\xi_\alpha$; and in lower panel (b) divided by $\sqrt \xi_\alpha$.
Figure 3 then tests whether a relationship $\gamma/\Omega_\alpha \sim \xi_\alpha$ (upper panel (a)) or $\gamma/\Omega_\alpha \sim \sqrt \xi_\alpha$ (lower panel (b)) holds in the linear phase. We can see that as expected \cite{Carbajal2014}, the growth rates of  modes $\omega = 7 \Omega_\alpha$ to $12\Omega_\alpha$ scale as $\sqrt \xi_\alpha$. Mode $\omega = 6 \Omega_\alpha$ does not follow this scaling and here marks the transition to the linearly stable regime.

The nonlinear simulations presented here resolve a longstanding question in the physics of fusion alpha particle confinement and stability in MCF plasmas. They confirm the MCI as the likely emission mechanism and greatly strengthen the basis for diagnostic exploitation of ICE in future burning plasmas.

 Our results suggest that a fast linear phase of the MCI deposits energy into the wave fields of all of the linearly unstable modes, and preferentially into the fastest growing modes $8 \leq l \leq 10$. It is these modes that dominate the power spectrum at the start of the nonlinear phase. Thus the system enters its nonlinear phase, which begins when the fastest growing mode saturates, imprinted with: first, the energy acquired by each individual mode according to its particular growth rate; and second, the total energy transferred from the energetic ion population to the excited fields, which scales with $n_\alpha/n_i$. During the nonlinear phase this energy is redistributed across all the harmonics $1 \leq l \leq 12$. The energy density of the fundamental and low cyclotron harmonic modes ($1 \leq l \leq 6$), which are linearly stable, are driven up by nonlinear coupling. In the saturated nonlinear regime, wave energy density scales approximately linearly with $n_\alpha$ at each harmonic, regardless of its linear stability characteristics.
  Our simulations then suggest that, for the range of $n_\alpha/n_i$ considered, the relative amplitudes of the cyclotron harmonic peaks of the saturated power spectra are independent of $n_\alpha/n_i$ to within error bars. The linearly stable, but nonlinearly excited low cyclotron harmonic modes ($1 \leq l \leq 6$) have peak intensity scaling that follows that established experimentally. The diagnostic exploitation of ICE in future deuterium-tritium plasmas rests on the empirical scaling between ICE signal intensity and fusion product concentration. The results presented here provide the first-principles physics basis for this observed scaling.

\begin{acknowledgments}
L.C. acknowledges the Mexican Council of Science and Technology (CONACyT) for support.
This work was part-funded by the RCUK Energy Programme [under grant EP/I501045] and the European Communities.
We acknowledge the EPSRC for financial support.
This work has been carried out within the framework of the EUROfusion Consortium and has received funding from the Euratom research and training programme 2014-2018 under grant agreement No 633053. The views and opinions expressed herein do not necessarily reflect those of the European Commission.
\end{acknowledgments}


\begin{thebibliography}{20}
\bibitem{McClements2015}  K. G. McClements, R. D'Inca, R. O. Dendy, L. Carbajal, S. C. Chapman, J. W. S. Cook, R. Harvey, B. Heidbrink, and S. Pinches, Nucl. Fusion, 55, 043013, (2015) 
   
\bibitem{X3} S. Cauffman, R. Majeski, K. G. McClements, and R. O. Dendy, Nucl. Fusion 35 1597 (1995)

\bibitem{X2a} G. A. Cottrell and R. O. Dendy, Phys. Rev. Lett. 60, 33 (1988)

\bibitem{Cottrell1993} G. A. Cottrell, V. P. Bhatnagar, O. Da Costa, R. O. Dendy, J. Jacquinot, K. G. McClements, D. C. McCune, M. F. F. Nave, P. Smeulders, and D. F. H. Start, Nucl. Fusion, 33, 1365, (1993)

\bibitem{Dendy1995} R. O. Dendy, K. G. McClements, C. N. Lashmore-Davies, G. A. Cottrell, R. Majeski, and S. Cauffman, Nucl. Fusion, 35, 1733, (1995)

\bibitem{X2b} K. G. McClements, C. Hunt, R. O. Dendy, and G. A. Cottrell, Phys. Rev. Lett. 82, 2099 (1999)

\bibitem{X4b} V. S. Belikov and Y. I. Kolesnichenko, Sov. Phys. Tech. Phys. 20 1146 (1976)

\bibitem{X4a} R. O. Dendy, K. G. McClements et al., Phys. Plasmas 1 1918 (1994) 

\bibitem{McClements1996} K. G. McClements, R. O. Dendy, C. N. Lashmore--Davies, G. A.  Cottrell, S. Cauffman, and R. Majeski, Phys. Plasmas, 3, 543, (1996)

\bibitem{X4c} T. F{\"u}l{\"o}p and M. Lisak, Nucl. Fusion 38 761 (1998)

\bibitem{X5a} N. Gorelenkov and C. Z. Cheng, Nucl. Fusion 35 1743 (1995)

\bibitem{X5b} H. Smith and E. Verwichte, Plasma Phys. Control. Fusion 51 075001 (2009)
   
\bibitem{Cook} J. W. S. Cook, R. O. Dendy, and S. C. Chapman, Plasma Phys. Cont. Fusion, 55, 065003, (2013)

\bibitem{Carbajal2014} L. Carbajal, R. O. Dendy, S. C. Chapman, and J. W. S. Cook, Phys. Plasmas, 21, 012106, (2014) 

\bibitem{15} W. W. Heidbrink, M. E. Austin, R. K. Fisher, M. García-Muñoz, G. Matsunaga, G. R. McKee, R. A. Moyer, C. M. Muscatello, M. Okabayashi, D. C. Pace, K. Shinohara, W. M. Solomon, E. J. Strait, M. A. Van Zeeland, and Y. B. Zhu, Plasma Phys. Control. Fusion 53, 085028 (2011)

\bibitem{16} R. D'Inca, M. Garcia-Munoz  et al., Proc. 38th EPS Conf. Plasma Phys. 2012 P1.053

\bibitem{17} R. D'Inca, PhD thesis, Max Planck Institute for Plasma Physics 2014

\bibitem{18} M. Ichimura, H. Higaki, S. Kakimoto, Y. Yamaguchi, K. Nemoto, M. Katano, M. Ishikawa, S. Moriyama, and T. Suzuki, Nucl. Fusion 48, 035012 (2008)

\bibitem{19} S. Sato, M. Ichimura, Y. Yamaguchi, M. Katano, Y. Imai, T. Murakami, Y. Miyake, T. Yokoyama, S. Moriyama, T. Kobayashi, A. Kojima, K. Shinohara, Y. Sakamoto, T. Watanabe, H. Hojo, and T. Imai, Plasma and Fusion Research 5, S2067 (2010)

\bibitem{20} R. O. Dendy, K. G. McClements, C. N. Lashmore-Davies, R. Majeski, and S. Cauffman, Phys. Plasmas 1, 3407 (1994)

\bibitem{21} K. Saito, R. Kumazawa et al., Plasma Sci. Technol. 15, 209 (2013)

\bibitem{21a} S. G. Thatipamula, G. S. Yun, J. Leem, H. K. Park, K. W. Kim, T. Akiyama, and S. G. Lee, Plasma Phys. Control. Fusion, accepted (2016)

\bibitem{McClements1993}K. G. McClements and R. O. Dendy, J. Geophys. Res., 98, 11689, (1993)

\bibitem{McClementsJGR}K. G. McClements, R. O. Dendy,  and C. N. Lashmore-Davies, J. Geophys. Res., 99, 685, (1994)
  
\bibitem{Rekaa} V. L. Rekaa, S. C. Chapman, and R. O. Dendy, Astrophys. J. , 791, 26, (2014)

\bibitem{X2c} J. W. S. Cook, S. C. Chapman, and R. O. Dendy, Phys. Rev. Lett. 105, 255003 (2010)

\bibitem{PROMETHEUS} PROMETHEUS++ code,  \url{http://www2.warwick.ac.uk/fac/sci/physics/research/cfsa/people/carbajal_gomez/plasma_modelling}
 
 \bibitem{Gingell2012} P. W. Gingell, S. C. Chapman, R. O. Dendy, and C. S. Brady, Plasma Phys.  Cont. Fusion, 54, 065005, (2012)

\bibitem{Gingell2014} P. W.  Gingell, S. C. Chapman, and R. O. Dendy, Plasma Phys.  Cont. Fusion, 56, 035012, (2014)
  
 \bibitem{Sydora1999} R. D. Sydora, Journal of Computational and Applied Mathematics, 109, 243, (1999)

\end{thebibliography}
\end{document}